\documentstyle[12pt]{article}

\begin{document}
\centerline {\Large \bf Controlling the Size of Popcorn}
\vskip 1.0 true cm
\centerline{\bf Daniel C. Hong and Joseph A. Both}
\vskip 0.2 true cm
\centerline{Physics, Lewis Laboratory, Lehigh University, Bethlehem,
Pennsylvaian 18015}
\date{\today}

\begin{abstract}
We present a thermo-statistical model of popcorn
production
and propose a way to control the final size of the popcorn by monitoring
only the chamber pressure.

\end{abstract}
\vskip 0.3 true cm
\noindent P.A.C.S: 05.20-y, 01.55.+b, 01.90+g, 05.70-2.


\vskip 0.3 true cm

\vskip 1.0 true cm
How can one control the size of popcorn?  The food industry may have
ample reason to find ways to double or triple the size of popcorn,
because large popcorn may be pleasing to the
eyes, possibly energy saving in its mass production,
and thus profitable to producers and consumers alike.  The purpose of this
note is to examine the physics of popcorn production
based on a few thermodynamic principles and recommend
ways to control the size of the final product.
Before going into detail, let us first examine the underlying physics of
popcorn production.  First, there is moisture inside the corn, which,
upon being heated in a chamber (e.g. a microwave oven), is subject to
expansion.
When the temperature
of the chamber exceeds the
boiling temperature, the moisture turns into a gas with a sudden expansion
in its volume.
But because of the hard shell that surrounds the corn, the moisture
is trapped
inside and its pressure rises. When the gas pressure
reaches the yield point of the shell,
the shell is broken and the trapped gas undergoes a rapid adiabatic
expansion, because the process is so fast that the gas does not have
enough time
to exchange heat and equilibrate itself with the environment. This
adiabatic expansion
stops when the gas pressure $P$ reaches that of the chamber,
$P_o$, which is, under normal
circumstances, the atmospheric pressure.  Since
our primary interest here is to control the final size of the
popcorn, and we know when the expansion stops, all we need do is
write down the governing equations of the interface right after the rupture of
the shell.
Hence, this problem is reduced to
that of interfacial instability and pattern formation, i.e.
we are dealing with an interface advancing into the chamber.

To write the equations, we recognize that the dynamic variable is obviously the
pressure $P$ inside the corn at time $t$, which should obey the law of
adiabatic expansion:
$$P(t)V(t)^{\gamma} = C_o = const \eqno (1)$$
where $V(t)$ is the volume of the corn, and $\gamma$ is the ratio of the
specific heats at constant pressure and at constant volume.
For water vapor, $\gamma\approx 1.3$ ({\it 1}).  The
constant $C_o$ is determined by the yield pressure, $P_Y$, of the shell
and the initial volume of the corn, $V_o$, i.e.
$C_o = P_Y V_o^{\gamma}.$
It is physical to assume that the advancing velocity of
the interface, or the normal velocity, $v_n$,
is proportional to the pressure gradient:
$$v_n= \kappa(P-P_o) \eqno (2)$$
where $\kappa $ is a material constant.  Eqs. (1) and (2) define the
dynamics of the interface.

We next examine the stability of the interface in a standard way.
We first scale the time in such a way that $\kappa =1$.
If the radius of the corn at time $t$ is
$R(t)$, then its
volume is simply given by $V(t)=\frac{4}{3}\pi R^3(t)$.
The evolution of the sphere is then governed by Eq. (2):
$$ v_n =dR(t)/dt = P-P_o = A/R^{3\gamma} - P_o \eqno (3)$$
where $A=C_o(\frac{3}{4\pi})^{\gamma}$.
Note that $\gamma$ is a non-integer, and thus
Eq. (3) is highly nonlinear, and the exact solution is not available.
However, the asymptotic form of the solution is easy to find.  In the
beginning, one may set $P_o = 0$, because $P_o<<P$,
and obtain an approximate solution,
$$R(t)^{1+3\gamma} \approx A(1+3\gamma)t + B \eqno (4)$$
where $B=R_o^{1+3\gamma}$ with
$R_o$ being the initial radius at $t=0$.  In the limit, $t\rightarrow\infty$,
the asymptotic solution is obtained by
setting the left hand side of Eq. (3) to zero, i.e.: $R(\infty)=
(A/P_o)^{1/3\gamma}.$  So, $R(t)$ initially increases as a power law, and then
approaches $R(\infty)$.  We have solved Eq. (3) numerically, and indeed
checked this behavior.
To examine stability of this solution, we imagine the sphere is perturbed
slightly.
Since any perturbations in three dimensions around the
sphere can be expanded in spherical harmonics, $Y_{lm}$,
let
$$ r(t,\theta,\phi) = R(t) + A_{lm}(t)Y_{lm}(\theta,\phi) \eqno (5)$$
If the amplitude $A_{lm}$ grows in time, then the interface is
unstable to the perturbation.  Otherwise, it is stable.
We should now obtain the corrections to the pressure and volume
caused by this perturbation.  Define $ \bar P(t) = P(t) + f(t)$ and
$ \bar V(t) = V(t) + g(t)$, where
$f$ and $g$ will be functions of $A_{lm}$ and $Y_{lm}$.  We first obtain
$g(t)$ in the case $l\ne 0$
by direct integration of the volume:
$ \bar V(t) = \frac{1}{3}\int r^3(t)d\Omega = \frac{4}{3}\pi R^3
+ RA^2_{lm}$. Hence, we find: $ g(t) = R A^2_{lm}$ and
$$ f(t)=-\frac{\gamma C_o}{V^{\gamma+1}}R A_{lm}^2 \eqno (6)$$
Now, in order to investigate the stability, we need a time
dependent equation of motion for the amplitude $A_{lm}$.  The normal
velocity, $v_n$, is given by:
$$ v_n = d{\bf r}/dt\bullet {\bf n} = P - P_o + f(t)\eqno (7)$$
where {\bf r} = $(x,y,z)=r(t,\theta,\phi)\left(\sin\theta \cos\phi, \sin\theta
 \sin\phi,\cos\theta\right)$ with $r(t,\theta,\phi)$ given by (6) and
{\bf n} is the normal vector to the surface, which is given by:
$$ {\bf n}=(-\partial z/\partial x, -\partial z/\partial y, 1)/\sqrt{1+(\partial
 z/\partial x)^2+(\partial z/\partial y)^2}\eqno (8)$$
Canceling $\dot R(t)=P-P_o$ (Eq. (3)), and calculating the derivatives by
chain rules yields:

$$ f(t) = \dot A_{lm} Y_{lm}(\theta, \phi)
-\frac{(\dot R + \dot A_{lm} Y_{lm}(\theta,\phi)) ((\partial_{\theta}Y_{lm})^2
\sin^2\theta + (\partial_{\phi}Y_{lm})^2)}{2R^2\sin^2\theta}A^2_{lm}
+ ...
\eqno (9)$$
Since there is no angular dependence in $f(t)$, we conclude from Eq. (9)
that $\dot A_{lm}=0$, namely
all higher order harmonic perturbations of $l \ne 0$
must be marginally stable.  If
we consider only the radial perturbation, then we may set,
$r(t)=R(t) + \delta(t)$.  Then it is easy to show that the growth rate
$\omega$ is negative:
$$\omega= \frac{\dot\delta}{\delta} = -\frac{4\pi R^2 C_o}{V^{\gamma+1}}
+ O(\delta ^2)< 0\eqno
(10)$$
Hence, the interface is stable against the radial
perturbation.
In summary, we have shown at the
level of the linear stability analysis, that perturbations of high order
harmonics are marginally stable, and the lowest
radial mode of $l=0$ decays exponentially.
We now present a physical argument why the interface is indeed
stable.  Note that the volume correction due to the
perturbation, $g(t)$, is positive, while the correction to the pressure, $f(t)$
 is
negative.
Thus, when the perturbation arises,
the inside  pressure that drives the instability decreases,
and the propagating speed decreases.  Consequently,
the
instability is suppressed.  Note that
this problem is analogous to the problem of a solidifying
interface
advancing from a cold to a hot environment,
which is always stable against perturbations.

Since we have established
that the interface is stable against infinitesimal perturbations,
we are now in a position to determine the approximate size
of the popcorn at a given chamber pressure $P_o$.
Suppose the corn stops its
expansion at a certain time $t_f$, at which point $P(t_f)=P_o$.
We can easily find from Eq. (1)
the maximum volume of the corn at $t_f$:
$$ V(t_f) = (C_o/P_o)^{1/\gamma} \eqno (11)$$
Note that $V(t_f)$ is
a function of the initial
chamber pressure $P_o$,
yield pressure $P_Y$ and the initial volume, $V_o$, because
$C_o=P_Y V_o^{\gamma}$.
For a given corn that has fixed $V_o$ and $P_Y$, Eq. (11) enables us
to control the size of the popcorn by
monitoring only the chamber pressure $P_o$.  Also, from Eq. (11), we
may have a rough estimate of the yield pressure.  Examining typical
popcorn, we note that the radius increases at least by a factor 4,
and thus $V_f/V_o \approx 60$.  Hence,
$$P_Y/P_o=(V_f/V_o)^{\gamma} \approx 200 $$
We note that the
yield strength of the shell at the rupture point
is about the same as or greater than
polyethylene (LDPE) at room temperature ({\it 2}).
We now
define the volume expansion rate $\Gamma$ as
the ratio of the final volume over the initial volume $V_o$:
$$ \Gamma(P_o) = V(t_f)/V_o = (C_o/P_o)^{1/\gamma}/V_o \eqno (12) $$
If our goal is simply to increase or decrease
the size of the popcorn relative to a given reference
point, then the more relevant quantity is
the ratio $\alpha \equiv \Gamma(P'_o)/\Gamma(P_o)$.  From Eq. (12), we obtain:
$$ \alpha = (P_o/P_o')^{1/\gamma} \eqno (13a)$$
or equivalently,
$$ P'_o= \alpha ^{-\gamma} P_o \eqno (13b)$$
which is the central result of this paper.
If we want to increase the size of the popcorn by a factor $\alpha$, then
we must reduce the pressure of the chamber by a factor $\alpha^{\gamma}$.
If $\alpha=2$, one must reduce the
chamber pressure by
a factor $2^{1.3} \approx 2.5$.
However, we caution that the assumption of our model that the solid expansion
of the corn via eversion
closely follows the adiabatic expansion of the gas may be
too simplistic, and thus the exponent, $1/\gamma$,
in (13a), which characterizes the expansion rate,
may be somewhat smaller
and presumably saturate
beyond the critical pressure.  We will report
the experimental results in the near future.
Another way of controlling the size of the popcorn
may be to increase the
pressure $P$ inside the corn and the chamber pressure $P_o$ simultaneously,
and let the chamber undergo a very rapid free expansion,
when the temperature of the chamber reaches the popping temperature.
During this
free expansion, the
chamber pressure  decreases, while the pressure inside remains constant.
When the difference, $P-P_o$,
reaches the yield point, $P_Y$, the shell breaks, and the
the corn undergoes the adiabatic expansion.  By controlling
the duration of the heating and the free expansion, one may
control the size of the final product.  This method may be in fact more
promising than the previous one, because by not reducing the
chamber pressure, it prevents moisture from being leaked to the chmaber
during heating, which in turn increases the pressure of the corn,
and hence the pressure gradient of the solid-gas interface during the
expansion.

\vskip 1.0 true cm
{\bf Acknowledgement}
\vskip 0.2 true cm
This work was inspired by a dinner speech delivered
by Dr. Young Hwa Kim of Higher Dimension Research
at the 2000 annual meeting of Association of Korean American Physicists.
DCH wishes to
thank him for bringing his attention to this problem,
as well as J. A. McLennan and H. D. Ou-Yang for helpful discussions.
\vskip 1.0 true cm

\newpage
\noindent{\bf References}
\vskip 1.0 true cm

\noindent 1.  {\it CRC Handbook of Physics and Chemistry} (CRC Press, Florida,
1985), p. D-172.

\noindent 2.  N. E. Dowling, {\it Mechanical Behavior of Materials}
(Prentice Hall, New Jersey, 1993), p. 158.

\end{document}